\documentclass[dvips]{acta}
\usepackage{supertabular,lscape,epsfig}
\usepackage{amssymb}
\usepackage{amsmath}
\DeclareSymbolFont{ppa}{OT1}{ppl}{m}{it}
\DeclareMathSymbol{\vv}{\mathalpha}{ppa}{'166}

\newfont{\hb}{rphvb at 10pt}
\newfont{\hbo}{rphvbo at 10pt}
\newfont{\bitt}{rptmbi at 12pt}
\newfont{\bits}{rptmbi at 11pt}

\SetPages{17}{39}

\SetVol{60}{2010}

\begin{document}

\newcommand{\TabCapp}[2]{\begin{center}\parbox[t]{#1}{\centerline{
  \small {\spaceskip 2pt plus 1pt minus 1pt T a b l e}
  \refstepcounter{table}\thetable}
  \vskip2mm
  \centerline{\footnotesize #2}}
  \vskip3mm
\end{center}}

\newcommand{\TTabCap}[3]{\begin{center}\parbox[t]{#1}{\centerline{
  \small {\spaceskip 2pt plus 1pt minus 1pt T a b l e}
  \refstepcounter{table}\thetable}
  \vskip2mm
  \centerline{\footnotesize #2}
  \centerline{\footnotesize #3}}
  \vskip1mm
\end{center}}

\newcommand{\MakeTableSepp}[4]{\begin{table}[p]\TabCapp{#2}{#3}
  \begin{center} \TableFont \begin{tabular}{#1} #4 
  \end{tabular}\end{center}\end{table}}

\newcommand{\MakeTableee}[4]{\begin{table}[htb]\TabCapp{#2}{#3}
  \begin{center} \TableFont \begin{tabular}{#1} #4
  \end{tabular}\end{center}\end{table}}

\newcommand{\MakeTablee}[5]{\begin{table}[htb]\TTabCap{#2}{#3}{#4}
  \begin{center} \TableFont \begin{tabular}{#1} #5 
  \end{tabular}\end{center}\end{table}}

\newfont{\bb}{ptmbi8t at 12pt}
\newfont{\bbb}{cmbxti10}
\newfont{\bbbb}{cmbxti10 at 9pt}
\newcommand{\uprule}{\rule{0pt}{2.5ex}}
\newcommand{\douprule}{\rule[-2ex]{0pt}{4.5ex}}
\newcommand{\dorule}{\rule[-2ex]{0pt}{2ex}}
\def\thefootnote{\fnsymbol{footnote}}
\hyphenation{Ce-phe-ids}
\begin{Titlepage}
\Title{The Optical Gravitational Lensing Experiment.\\
The OGLE-III Catalog of Variable Stars.\\
VII.~Classical Cepheids in the Small Magellanic Cloud\footnote{Based on
observations obtained with the 1.3-m Warsaw telescope at the Las Campanas
Observatory of the Carnegie Institution of Washington.}}
\Author{I.~~S~o~s~z~y~ñ~s~k~i$^1$,~~
R.~~P~o~l~e~s~k~i$^1$,~~
A.~~U~d~a~l~s~k~i$^1$,\\
M.\,K.~~S~z~y~m~a~ñ~s~k~i$^1$,~~
M.~~K~u~b~i~a~k$^1$,~~
G.~~P~i~e~t~r~z~y~ñ~s~k~i$^{1,2}$,\\
£.~~W~y~r~z~y~k~o~w~s~k~i$^3$,~~
O.~~S~z~e~w~c~z~y~k$^2$,~~
and~~K.~~U~l~a~c~z~y~k$^1$}
{$^1$Warsaw University Observatory, Al.~Ujazdowskie~4, 00-478~Warszawa, Poland\\
e-mail:
(soszynsk,rpoleski,udalski,msz,mk,pietrzyn,kulaczyk)
@astrouw.edu.pl\\
$^2$ Universidad de Concepci{\'o}n, Departamento de Astronomia, Casilla 160--C,
Concepci{\'o}n, Chile\\
e-mail: szewczyk@astro-udec.cl\\
$^3$ Institute of Astronomy, University of
Cambridge, Madingley Road, Cambridge CB3 0HA, UK\\
e-mail: wyrzykow@ast.cam.ac.uk}
\Received{March 2, 2010}
\end{Titlepage}
\Abstract{The seventh part of the OGLE-III Catalog of Variable Stars
(OIII-CVS) consists of 4630 classical Cepheids in the Small Magellanic
Cloud (SMC). The sample includes 2626 fundamental-mode (F), 1644
first-overtone (1O), 83 second-overtone (2O), 59 double-mode F/1O, 215
double-mode 1O/2O, and three triple-mode classical Cepheids. For each object
basic parameters, multi-epoch {\it VI} photometry collected within 8 or 13
years of observations, and finding charts are provided in the OGLE Internet
archive.

We present objects of particular interest: exceptionally numerous sample of
single-mode second-overtone pulsators, five double Cepheids, two Cepheids
with eclipsing variations superimposed on the pulsation light curves. At
least 139 first-overtone Cepheids exhibit low-amplitude secondary
variations with periods in the range 0.60--0.65 of the primary ones. These
stars populate three distinct sequences in the Petersen diagram. The origin
of this secondary modulation is still unknown. Contrary to the Large
Magellanic Cloud (LMC) we found only a few candidates for anomalous
Cepheids in the SMC. This fact may be a clue for the explanation of the
origin of the anomalous Cepheids. The period and luminosity distributions
of Cepheids in both Magellanic Clouds suggest that there are two or three
populations of classical Cepheids in each of the galaxies. The main
difference between the LMC and SMC lays in different numbers of Cepheids in
each group. We fit the period--luminosity (PL) relations of SMC Cepheids
and compare them with the LMC PL laws.}{Cepheids -- Stars: oscillations --
Magellanic Clouds}

\Section{Introduction}
Classical Cepheids (also called $\delta$~Cep stars, type I Cepheids or
Population I Cepheids) provide us with precise tests of stellar interiors
and evolution. These are also the milestones of the extragalactic distance
scale through their famous period--luminosity (PL) relation. Huge samples
of Cepheids and other variable stars detected in recent years by
microlensing surveys have increased our knowledge about statistical
features of these objects, but also reveal very rare or previously unknown
subtypes of variable stars.

Classical Cepheids in the Small Magellanic Cloud (SMC) played an important
historical role. Leavitt (1908) derived periods for 16 SMC Cepheids and
noticed that ``the brighter variables have the longer periods'', which was
the first hint of the PL relation. Within the next few decades the number of
known Cepheids in the SMC grew significantly, mainly as a result of the
extensive survey of the Magellanic Clouds carried out by the Harvard
Observatory. In 1955 periods were derived for 670 Cepheids in the SMC
(Shapley and McKibben Nail 1955, and references therein). The Harvard
catalog of variable stars in the SMC by Payne-Gaposchkin and Gaposchkin
(1966) comprised as many as 1155 Cepheids.

A larger sample of Cepheid variables in the SMC was only released when the
Optical Gravitational Lensing Experiment entered its second phase (OGLE-II)
and collected the photometric database of stars in the Magellanic
Clouds. Udalski \etal (1999abd) published the catalog of Cepheids in the
SMC, consisting of 2062 single-mode pulsators (including several type~II
Cepheids) and 93 double-mode variables. From the recent catalogs, it is
worth mentioning the EROS-2 catalog of double-mode Cepheids in the
Magellanic Clouds (Marquette \etal 2009), which includes 170 beat Cepheids
in the SMC.

The OGLE-III Catalog of Variable Stars (OIII-CVS) is intended to include
all variable sources detected among about 400~million stars monitored
during the third phase of the OGLE survey. In the previous papers of this
series we presented the catalogs of over 120\,000 pulsating stars in the
Large Magellanic Cloud (LMC): classical Cepheids (Soszyñski \etal 2008b,
hereafter Paper~I), type~II and anomalous Cepheids (Soszyñski \etal 2008c),
RR~Lyr stars (Soszyñski \etal 2009a), long-period variables (Soszyñski
\etal 2009b), and $\delta$~Sct stars (Poleski \etal 2010). In the present
paper we describe the first part of the OIII-CVS that contains variable
stars in the SMC.

Our sample doubles the number of known classical Cepheids in the SMC. The
catalog comprises 4630 variables -- the largest set of Cepheids identified
to date in this and any other galaxy. As in previous parts of the OIII-CVS,
we release the photometric and astrometric information about each
star. These data are described in Section~2. In Section~3 we present
methods used to select Cepheids in the SMC. The catalog itself is described
in Section~4. In Section~5 we compare the distributions of periods and
luminosities of the LMC and SMC Cepheids. In Section~6 we present the PL
relations and in Section~7 we draw our conclusions.

\Section{Observational Data}
The observational data provided with this catalog were collected during the
OGLE-III project between 2001 and 2009 with the 1.3-meter Warsaw telescope
at Las Campanas Observatory, Chile. The observatory is operated by the
Carnegie Institution of Washington. The telescope was equipped with the
``second generation'' camera consisting of eight SITe $2048\times4096$ CCD
detectors with 15~$\mu$m pixels which corresponded to 0.26 arcsec/pixel
scale. The gain of the chips was adjusted to be about
1.3~$\mathrm{e}^-/\mathrm{ADU}$ with the readout noise from 6 to 9
$\mathrm{e}^-$ depending on the chip. Details of the instrumentation setup
can be found in Udalski (2003).

Observations covered over 14 square degrees (41 fields) distributed over
the densest regions of the SMC. Typically about 700 points in the Cousins
{\it I}-band and 50--70 points in the Johnson {\it V}-band were collected,
although for some stars we obtained two or even three times more points,
since they were located in the overlapping parts of two or three adjacent
fields. It is worth noting that imperfections in the telescope pointing
compensated for small gaps between the CCD chips of the mosaic, so the
completeness of the catalog is virtually not affected by these gaps, though
stars located at the very edge of the fields sometimes have a smaller number
of observations.

For about 2300 stars in the central 2.4 square degrees of the SMC the
OGLE-III photometry was supplemented with the OGLE-II observations
(Szymañski 2005) collected from 1997 through 2000. These are additional
300--400 points in the {\it I}-band and 30--40 measurements in the {\it
V}-band. We tied both datasets by shifting the OGLE-II photometry to agree
with the OGLE-III light curves.

The photometry was obtained with the standard OGLE data reduction pipeline
(Udalski \etal 2008) based on the Difference Image Analysis (DIA, Alard and
Lupton 1998, Alard 2000, Wo¼niak 2000). For 17 brightest Cepheids there is
no DIA {\it I}-band photometry, because these stars saturate in the
reference frames. Instead, we provide the {\sc DoPhot} (Schechter \etal
1993) photometry for these objects, derived independently on individual
images. These objects are flagged in the remarks of the catalog. Since at
least part of the observations of these bright objects were saturated, the
{\sc DoPhot} light curves are much more noisy than the DIA ones, but their
quality is good enough to study periods and light curve
shapes. Nevertheless, we do not recommend using this photometry for
absolute calibration of the Cepheid brightness.

\Section{Selection of Cepheids}
\Subsection{Single-Mode Cepheids}
The process of the variable stars selection began with a period search for
all stars monitored during the OGLE-III project in the SMC. About 6 million
stars were subjected to a Fourier-based frequency analysis with the {\sc
Fnpeaks} code\linebreak (Z.~Ko³aczkowski, private communication) at the
Interdisciplinary Centre for Mathematical and Computational Modelling of
the University of Warsaw (ICM). For each star the ten highest peaks in the
periodograms were selected and archived with the corresponding amplitudes
and signal-to-noise (S/N) ratios. Then, the light curve was prewhitened
with the primary period and the procedure of the period search was repeated
on the residual data.

For every object with ${\rm S/N}>5$ we derived the mean {\it I}- and {\it
V}-band magnitudes and constructed the period--$W_I$ diagram, where $W_I$
is the extinction-insensitive Wesenheit index, defined as:
$$W_I=I-1.55(V-I).$$ 
In this plane classical Cepheids follow distinct PL relations and were easy
to spot. For further analysis we selected stars located in the wide region
above and below these sequences in order to also include blended Cepheids
and type II Cepheids. All these light curves were visually inspected and
divided into three groups: pulsation-like, eclipsing-like and other
variable objects. The pulsation-like light curves were tentatively
categorized into classical Cepheids, type II Cepheids, RR~Lyr stars and
$\delta$~Sct stars. To distinguish between first-overtone Cepheids and
$\delta$~Sct stars we adopted the same limiting period as in the LMC \ie
0.24~days. We emphasize that this is an arbitrary value, because
$\delta$~Sct stars are located at the extension of the PL relation of
Cepheids, so our shortest-period Cepheids may be called $\delta$~Sct stars
in other investigations.

During the selection process we identified several dozen type II
Cepheids in the SMC. Their list will be published in the subsequent part of
the OIII-CVS. Somewhat surprisingly, we did not notice any PL relation
spreading between classical and type~II Cepheids and populated by anomalous
Cepheids, like in the LMC where 62 fundamental-mode anomalous Cepheids were
found (Soszyñski \etal 2008c). In the SMC we selected only 3 candidates for
fundamental-mode and 3 candidates for first-overtone anomalous
Cepheids. These stars are located slightly below the PL relations for
classical Cepheids and their light curves are similar to the light curves
of anomalous Cepheids in the LMC. Two of the candidate fundamental-mode
anomalous Cepheids have periods shorter than any other fundamental-mode
classical Cepheid. However, it cannot be excluded that these objects are
classical Cepheids located farther than other classical Cepheids or
are RR~Lyr stars in the foreground of the SMC. Table~1 gives basic
information about our six candidates for anomalous Cepheids in the
SMC. These stars are also flagged in the remarks of the Catalog.

\tabcolsep=3pt
\MakeTableee{lcccccc}{12.5cm}{Candidates for anomalous Cepheids in the SMC}
{\hline
\noalign{\vskip3pt}
\multicolumn{1}{c}{Star name} & RA & DEC & Pulsation & $P$ & $\langle{I}\rangle$ & $\langle{V}\rangle$ \\
& [J2000.0] & [J2000.0] & Mode & [days] & [mag] & [mag] \\
\noalign{\vskip3pt}
\hline
\noalign{\vskip3pt}
OGLE-SMC-CEP-0366 & 00\uph40\upm36\zdot\ups07 &
$-72\arcd38\arcm04\zdot\arcs2$ & 1O & 0.6209232 & 18.254 & 18.611 \\
OGLE-SMC-CEP-0677 & 00\uph44\upm42\zdot\ups95 &
$-73\arcd11\arcm26\zdot\arcs8$ & F & 0.8278529 & 18.460 & 19.167 \\
OGLE-SMC-CEP-2343 & 00\uph55\upm07\zdot\ups45 &
$-72\arcd44\arcm34\zdot\arcs1$ & 1O & 0.5703437 & 18.342 & 18.779 \\
OGLE-SMC-CEP-2740 & 00\uph57\upm46\zdot\ups17 &
$-72\arcd14\arcm33\zdot\arcs6$ & F & 0.8297559 & 18.249 & 18.835 \\
OGLE-SMC-CEP-3540 & 01\uph03\upm45\zdot\ups77 &
$-73\arcd53\arcm09\zdot\arcs8$ & 1O & 0.5208060 & 18.387 & 18.794 \\
OGLE-SMC-CEP-4608 & 01\uph22\upm07\zdot\ups39 &
$-71\arcd31\arcm47\zdot\arcs9$ & F & 1.2563281 & 17.819 & 18.256 \\
\noalign{\vskip3pt}
\hline}

\begin{figure}[htb]
\centerline{\includegraphics[width=12.4cm]{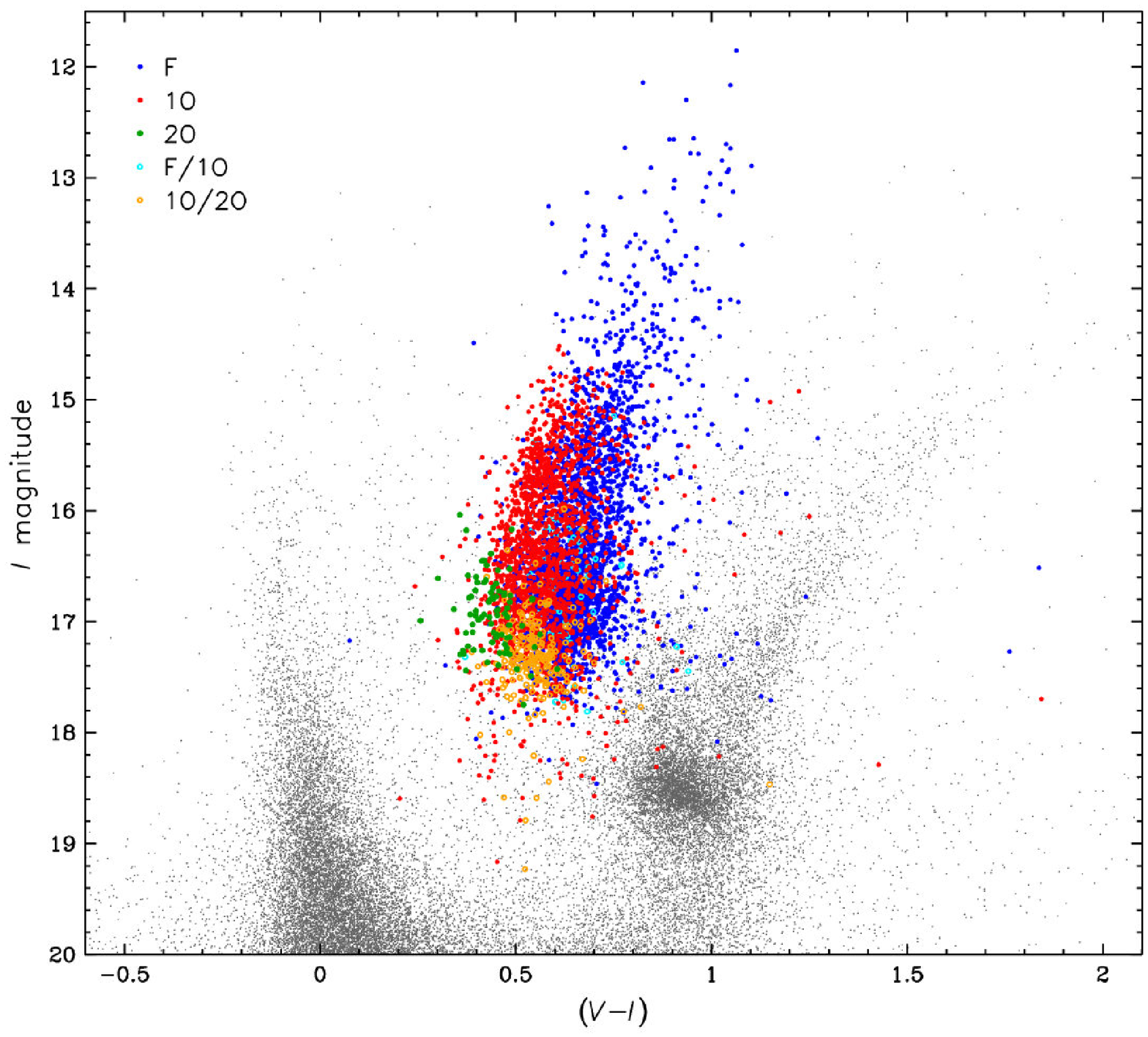}}
\FigCap{Color--magnitude diagram for classical Cepheids in the SMC. Blue
symbols show single-mode F Cepheids, red -- 1O, green -- 2O, cyan --
double-mode F/1O and orange -- 1O/2O pulsators. The background gray points
show stars from the subfield SMC100.1.}
\end{figure}
A modest number of anomalous Cepheids in our data was rather unexpected,
because the General Catalog of Variable Stars (GCVS, Artyukhina \etal 1995)
listed as many as 42 BL~Boo stars (\ie anomalous Cepheids) in the SMC. We
verified this classification for 38 variables that are located in the
OGLE-III fields. Most of them (31 objects) are regular classical Cepheids
populating the same PL relations as other pulsators of that type. For 17
stars from this group the periods provided in the GCVS are daily aliases of
the real periods. The remaining stars categorized as anomalous Cepheids in
the GCVS have periods below 1 day and are brighter than classical Cepheids
with similar periods. These are likely Galactic RR~Lyr stars in the
foreground of the SMC. It is worth noting that recently Bernard \etal
(2010) reported the absence of anomalous Cepheids in another metal-poor
galaxy -- IC~1613. This fact may indicate that anomalous Cepheids avoid
galaxies such as SMC or IC~1613, or that in these galaxies anomalous
Cepheids are indistinguishable from short-period classical Cepheids.

The final list of classical Cepheids was prepared after careful visual
inspection of the light curves, mean brightness and colors of the stars,
including near-infrared magnitudes from the IRSF/SIRIUS survey (Kato \etal
2007). We also utilized the information about proper motions of stars
archived in the OGLE databases to remove a few evident Galaxy members from
the list of candidates for SMC Cepheids. We believe the sample of Cepheids
published in this catalog is very clean, although it includes several
doubtful objects, \eg stars which obey the PL relations in each bandpass,
but show unusual light curves. Several objects show Cepheid-like light
curves, but they are located outside the instability strip in the
color--magnitude diagram (CMD, Fig.~1). The photometry of these stars may
be affected by a strong reddening or by crowding. We mark these objects as
``uncertain'' in the catalog remarks.
\begin{figure}[p]
\centerline{\includegraphics[width=13.0cm]{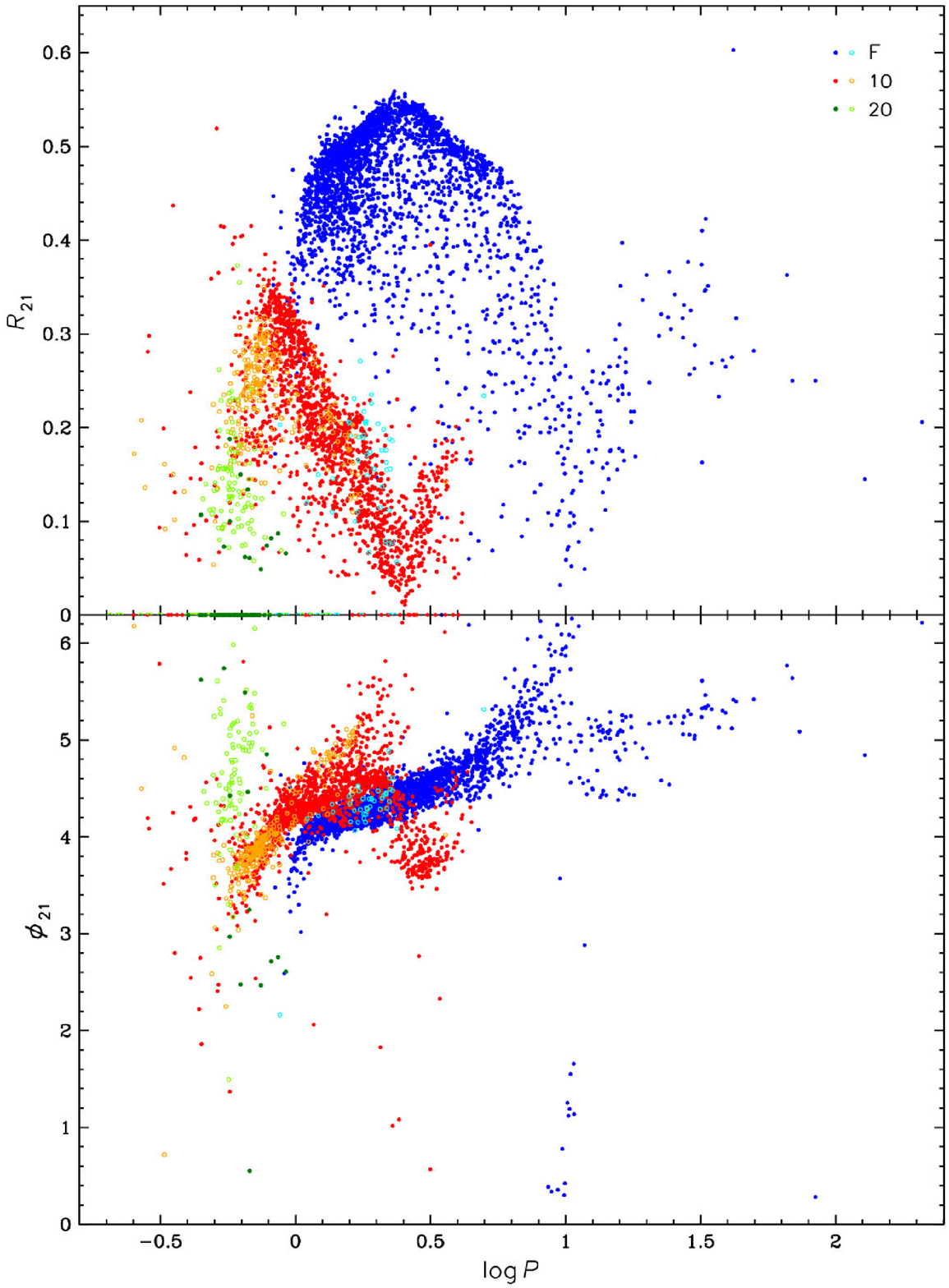}}
\FigCap{Fourier parameters $R_{21}$ and $\phi_{21}$ \vs $\log{P}$ for
classical Cepheids in the SMC. Blue, red, and dark-green solid circles mark
F, 1O and 2O single-mode Cepheids, respectively. Cyan, orange and
light-green empty circles represent F, 1O and 2O modes from double-mode
Cepheids (two points per star).}
\end{figure}
\begin{figure}[p]
\centerline{\includegraphics[width=13.0cm]{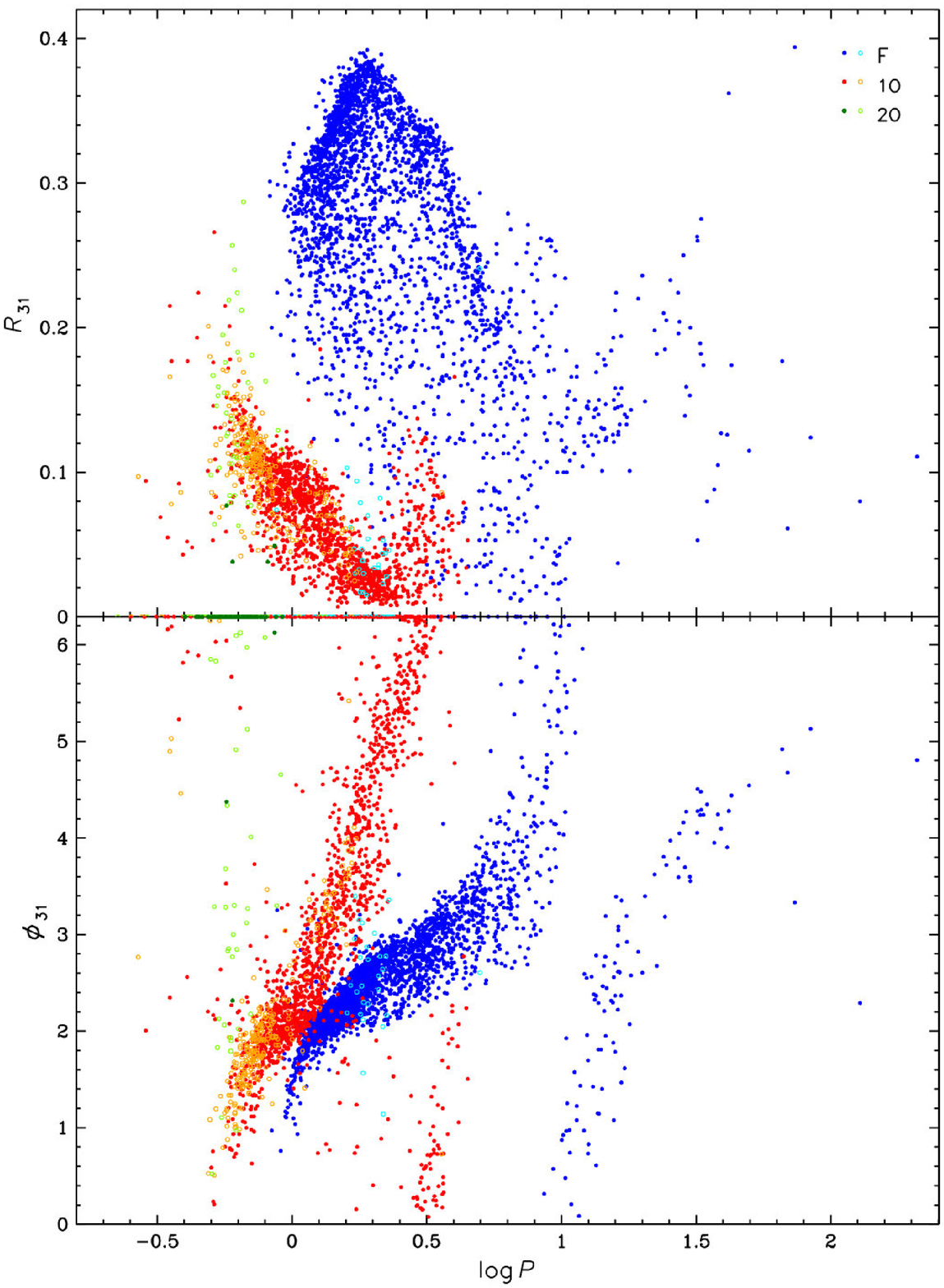}}
\FigCap{Fourier parameters $R_{31}$ and $\phi_{31}$ \vs $\log{P}$ for
classical Cepheids in the SMC. Color symbols represent the same modes of
pulsation as in Fig.~2.}
\end{figure}

The division into fundamental-mode (F) and first-overtone (1O) pulsators
was performed based on both, the position in the PL diagrams and the light
curve morphology represented by the Fourier coefficients. Figs.~2 and~3
show Fourier parameters $R_{21}$, $\phi_{21}$, $R_{31}$ and $\phi_{31}$
(Simon and Lee 1981) plotted as a function of $\log{P}$. When the amplitude
of the first (second) harmonic is statistically insignificant, then
$R_{21}=0$ ($R_{31}=0$) and $\phi_{21}$ ($\phi_{31}$) is undefined. Our
catalog includes several variables with atypical light curves that do not
match the patterns visible in Figs.~2 and~3. Their position in the PL
diagram does not always unambiguously indicate the mode of pulsation,
because of the overlap between PL sequences. Thus, our classification may
be uncertain for several single-mode Cepheids in the SMC.

We paid special attention to single-mode second-overtone (2O)
Cepheids. Udalski \etal (1999b) identified 13 such stars in the SMC, while
in the LMC we found 14 Cepheids pulsating solely in the second overtone
(Paper~I). As a result of our selection process in the SMC, we found a
considerable number of periodic stars with nearly sinusoidal light curves,
relatively small amplitudes ($A_I<0.14$~mag), located just above the PL
relation of 1O Cepheids and at the blue edge of the instability strip in
the CMD (Fig.~1). These stars have been recognized as 2O Cepheids. Our
catalog includes 83 firm candidates for 2O pulsators with periods in the
range of 0.40--0.92~days. This is, obviously, the largest sample of
single-periodic second overtone Cepheids detected so far.

The SMC hosts proportionally much larger number of 2O Cepheids (1.8\%) than
the LMC (0.4\%). Bernard \etal (2010) identified two second-overtone
pulsators (4\%) among 49 classical Cepheids in the IC~1613. This confirms
the suggestion of Antonello and Kanbur (1997) that 2O Cepheids are more
common in metal-poor environments.

\Subsection{Multimode Cepheids}
Multimode pulsating stars are very valuable objects, because each mode
gives independent constraints on stellar parameters (\eg Buchler and Szabó
2007, Dziembowski and Smolec 2009a). First double-mode Cepheids (called
also beat Cepheids) in the SMC were discovered by Beaulieu \etal (1997),
who reported the sample of 4 stars beating in the fundamental mode and
first overtone (F/1O) and 7 first/second overtone (1O/2O) pulsators. Then,
the list of double-mode Cepheids in the SMC was extended by Alcock \etal
(1997, 7 F/1O and 20 1O/2O Cepheids), Udalski \etal (1999a, 23 F/1O and 70
1O/2O Cepheids) and Marquette \etal (2009, 41 F/1O and 129 1O/2O Cepheids).

Double-mode Cepheids presented in this catalog were identified with two
methods. First, we used the periods derived for all stars observed in the
SMC to select objects with two primary periods typical for beat
Cepheids. The light curves of the tentatively selected objects were
inspected by eye and some of them were judged to be F/1O or 1O/2O
Cepheids. Second, we tested the secondary periods of all previously
identified Cepheids. Each light curve was fitted with a Fourier series with
a number of harmonics minimizing the $\chi^2$ per degree of freedom, the
function was subtracted from the observational data and the period search
was performed on the residual data. We visually examined the light curves
with characteristic period ratios and those with the largest S/N of the
secondary periods. This iterative procedure was repeated several times on
each light curve, to ensure that we found all double-mode Cepheids and to
identify possible triple-mode pulsators.

\begin{figure}[htb]
\centerline{\includegraphics[width=12.8cm, bb=25 265 565 745]{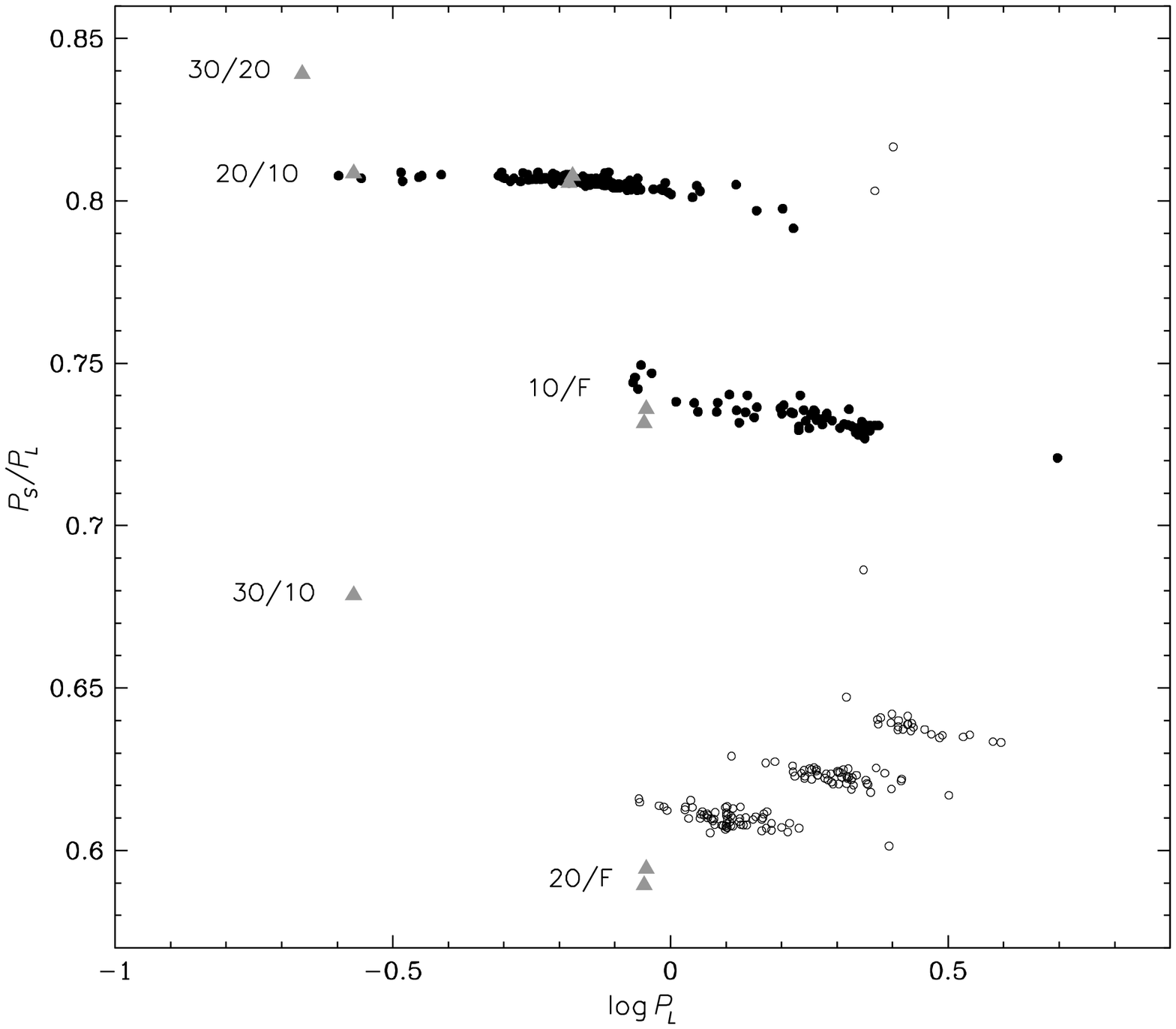}}
\FigCap{Petersen diagram for multiperiodic Cepheids in the SMC. Solid dots
represent double-mode (F/1O and 1O/2O) Cepheids, gray triangles show
triple-mode Cepheids (three points per star) and empty circles show
selected other stars with significant secondary periods.}
\end{figure}

Sometimes the Cepheids in our database presented very high rates of period
changes. In such cases the secondary periods were usually very close to the
primary ones. To search for beat Cepheids among these objects we performed
an additional procedure. For each Cepheid the photometry was divided into
individual observing seasons and the periods were adjusted separately for
each of these chunks. Then, these periods were used to prewhiten the light
curves independently in each season, and the frequency analysis was carried
out on the residual data. In this way we identified several additional
double-mode Cepheids.

In total, we detected 59 F/1O, 215 1O/2O and three triple-mode Cepheids in
the SMC. The Petersen diagram (shorter-to-longer period ratio \vs the
logarithm of the longer period) is shown in Fig.~4. The number of multimode
Cepheids in the SMC is almost the same as in the LMC (61 F/1O, 206 1O/2O, two
1O/3O and five triple-mode Cepheids), but, naturally, their incident rate
(6\%) is smaller than in the LMC (8\%), because the total number of
classical Cepheids in the SMC is larger than in the LMC. We draw the
reader's attention to the exceptionally long-period F/1O Cepheid
OGLE-SMC-CEP-1497, with the fundamental-mode period of about 5~days, but
connected with an extremely small (and thus uncertain) amplitude of about
0.008~mag. A similar outlying F/1O Cepheid (of even longer periods) was
found in the LMC (Paper~I). Another interesting object among double-mode
Cepheids is OGLE-SMC-CEP-0998 -- the only 1O/2O Cepheid in both Magellanic
Clouds with the second-overtone dominating over the first overtone.

\tabcolsep=2.5pt
\MakeTable{lcccccccc}{12.5cm}{Candidates for triple-mode Cepheids in the SMC}
{
\multicolumn{9}{c}{F/1O/2O Cepheids} \\
\noalign{\vskip1pt}
\hline
\noalign{\vskip3pt}
\multicolumn{1}{c}{Star name} & $\langle{I}\rangle$ & $\langle{V}\rangle$ & $P_F$ & $A_I^F$ & $P_{\rm 1O}$ & $A_I^{\rm 1O}$ & $P_{\rm 2O}$ & $A_I^{\rm 2O}$ \\
& [days] & [mag] & [days] & [mag] & [days] & [mag] & [mag] & [mag] \\
\noalign{\vskip3pt}
\hline
\noalign{\vskip3pt}
OGLE-SMC-CEP-1077 & 17.599 & 18.137 & 0.89734 & 0.018 & 0.6565184 & 0.301 & 0.528880 & 0.046 \\
OGLE-SMC-CEP-1350 & 17.471 & 18.087 & 0.90494 & 0.010 & 0.6659937 & 0.239 & 0.537878 & 0.057 \\
\noalign{\vskip2pt}
\hline
\noalign{\vskip15pt}
\multicolumn{9}{c}{1O/2O/3O Cepheid} \\
\noalign{\vskip1pt}
\hline
\noalign{\vskip3pt}
\multicolumn{1}{c}{Star name} & $\langle{I}\rangle$ & $\langle{V}\rangle$ & $P_{\rm 1O}$ & $A_I^{\rm 1O}$ & $P_{\rm 2O}$ & $A_I^{\rm 2O}$ & $P_{\rm 3O}$ & $A_I^{\rm 3O}$ \\
& [days] & [mag] & [days] & [mag] & [days] & [mag] & [mag] & [mag] \\
\noalign{\vskip3pt}
\hline
\noalign{\vskip3pt}
OGLE-SMC-CEP-3867 & 18.585 & 19.055 & 0.2688471 & 0.269 & 0.2173800 & 0.052 & 0.1824204 & 0.058 \\
\noalign{\vskip2pt}
\hline}

\begin{figure}[htb]
\centerline{\includegraphics[width=13.2cm]{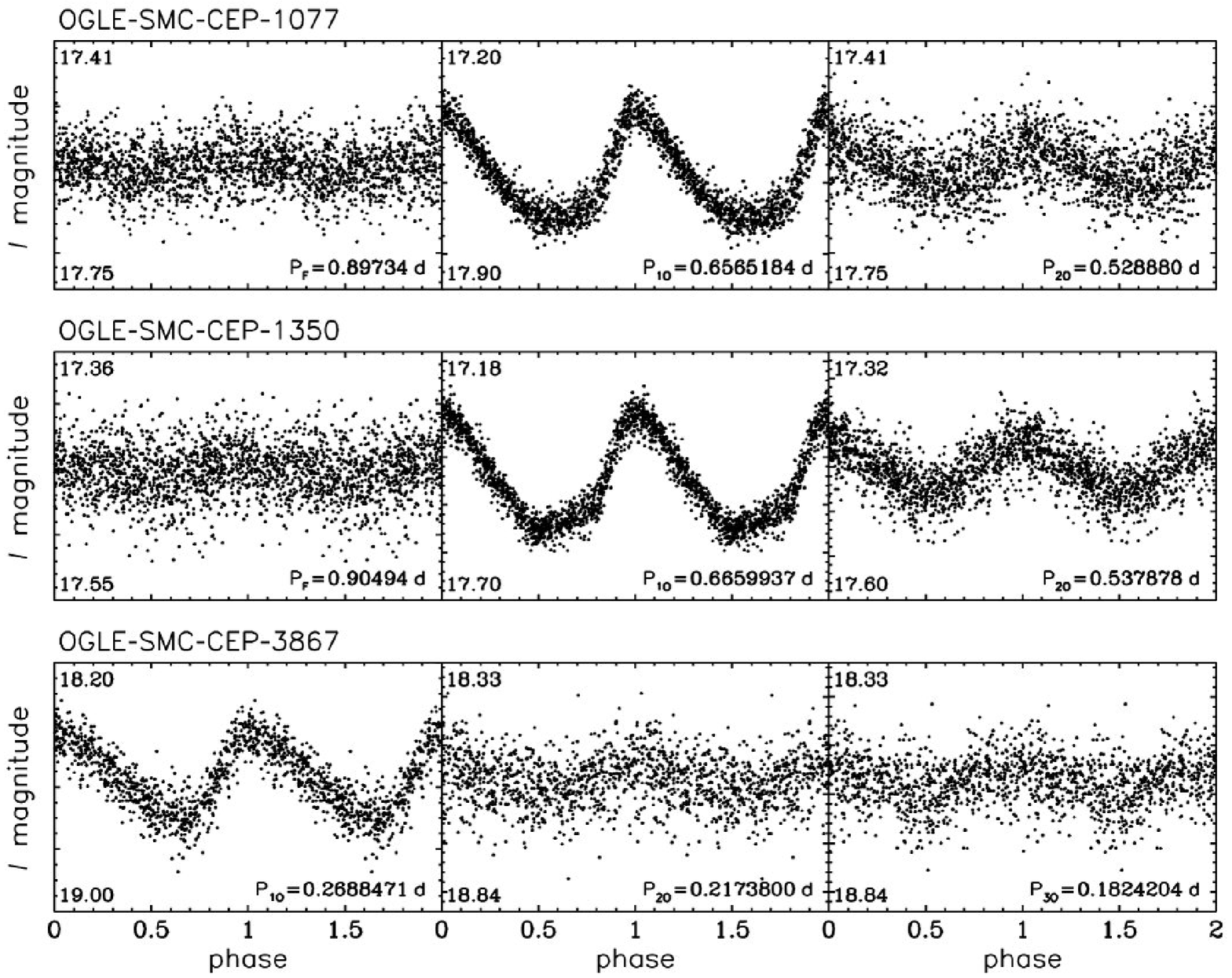}}
\FigCap{{\it I}-band light curves of triple-mode Cepheids in the SMC. 
Each mode is shown after prewhitening with the other two modes. Note that
the range of magnitudes varies from panel to panel. Numbers in the
left corners show the lower and the upper limits of magnitudes.}
\end{figure}

\tabcolsep=3pt
\MakeTable{lcccc}{12.5cm}{Double Cepheids in the SMC}
{\hline
\noalign{\vskip3pt}
\multicolumn{1}{c}{Star name} & Mode of & $P_1$ & Mode of & $P_2$ \\
 & Cepheid 1 & [days] & Cepheid 2 & [days] \\
\noalign{\vskip3pt}
\hline
\noalign{\vskip3pt}
OGLE-SMC-CEP-1526 & F & 1.2902286 & F & 1.8043107 \\
OGLE-SMC-CEP-2699 & F & 2.1174814 & 1O & 2.5623049 \\
OGLE-SMC-CEP-2893 & F & 1.3215528 & F & 1.1358572 \\
OGLE-SMC-CEP-3115 & F & 1.2519421 & F & 1.1597911 \\
OGLE-SMC-CEP-3674 & F & 2.8960473 & 1O & 1.8277592 \\
\noalign{\vskip3pt}
\hline}

\tabcolsep=3pt
\MakeTable{lccccc}{12.5cm}{Cepheids with additional eclipsing variability}
{\hline
\noalign{\vskip3pt}
\multicolumn{1}{c}{Star name} & Pulsation & $P_{\rm cep}$ & $P_{\rm ecl}$ &
$A_I^{\rm ecl}$ & $T_{\rm min}^{\rm ecl}$ \\
 & Mode & [days] & [days] & [mag] & [HJD] \\
\noalign{\vskip3pt}
\hline
\noalign{\vskip3pt}
OGLE-SMC-CEP-0411 & 1O & 1.1009833 & 43.4979 & 0.102 & 2450626.7133 \\
OGLE-SMC-CEP-1996 & 1O & 2.3179487 & 95.5944 & 0.018 & 2450424.4961 \\
\noalign{\vskip3pt}
\hline}

\begin{figure}[htb]
\centerline{\includegraphics[width=13cm, bb=55 455 565 745]{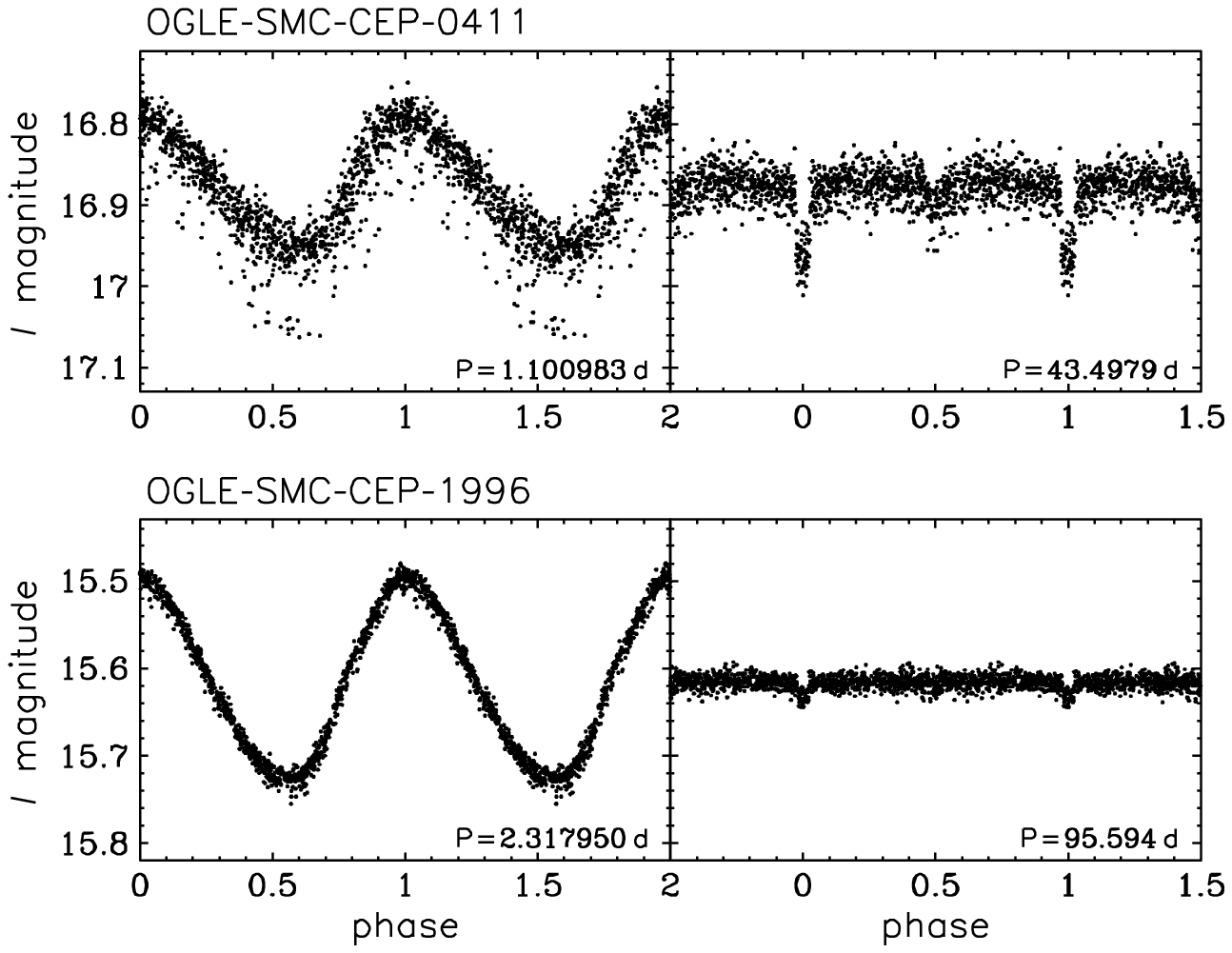}}
\FigCap{Light curves of Cepheids with additional eclipsing
variability. {\it Left panels} show the original photometric data folded
with the Cepheid periods. {\it Right panels} show the eclipsing light
curves after subtracting the Cepheid component. The ranges of magnitudes
are the same in each pair of the panels.}
\end{figure}

The main observational properties of the three candidates for triple-mode
Cepheids in the SMC are summarized in Table~2. Fig.~5 shows their light
curves folded with three primary periods, each one after prewhitening with
the other two modes. As one can see, the dominant periods of these stars
are always related to the first-overtone mode, just like in the five known
triple-mode Cepheids in the LMC (Soszyñski \etal 2008a). The mode
identification in our triple-mode Cepheids relies on their position in the
Petersen diagram (gray triangles in Fig.~4). Two stars have fundamental
mode, first and second overtones (F/1O/2O) excited, while one object is a
1O/2O/3O pulsator. The fundamental-mode variations in the former Cepheids
are of very low amplitudes, just above the detection limits of the OGLE
photometry, and should be treated as uncertain detections. The latter star
is undoubtedly a triple-mode pulsator, but the periods are very short, thus
it can also be classified as a $\delta$~Sct star. As we pointed out
already, the nomenclature is a matter of convention.

During our search for multiperiodicity we found a considerable number
(139~objects) of first-overtone Cepheids with additional small-amplitude
($A_I<0.015$~mag) modulation with periods of about 0.60--0.65 of the
dominant ones. These stars follow three distinct sequences in the Petersen
diagram (Fig.~4). The first stars of this kind were discovered in the LMC
by Moskalik and Ko³aczkowski (2008). In Paper~I we identified over 30 such
Cepheids in the LMC and noticed that these stars follow two sequences in
the Petersen diagram. In the SMC we identified three sequences and a much
bigger number of stars of that type, but the origin of these secondary
periods remains a mystery (Dziembowski and Smolec 2009b).

As a by-product of our analysis we found five double Cepheids -- unresolved
pairs of Cepheids which may be physically related. Only one object of that
type (OGLE-SMC-CEP-1526 = SMC\_SC5\_208044) was known before in the SMC
(Udalski \etal 1999a). Table~3 lists all double Cepheids in our
catalog. Information about these objects is also given in the remarks.

Two of our Cepheids show additional eclipsing variations superimposed on
the pulsational light curves. These objects may be optical blends of a
Cepheid and eclipsing variable, or the pulsating star may be a member of
the eclipsing binary system. Table~4 presents periods, amplitudes and
epochs of the minimum light of the eclipsing variability in these
potentially interesting objects. Fig.~6 presents the original photometry
folded with the Cepheid periods and the residual data (after subtracting
the pulsations) phased with the eclipsing periods. Note, the eclipses in
OGLE-SMC-CEP-1996 are very shallow, at the level of about 0.018~mag.

Finally, a significant number of Cepheids exhibit secondary periods very
close to the primary ones. After the first iteration of the multiperiod
analysis (\ie after prewhitening with the primary period), such a behavior
was observed for about 2\% of F, 33\% of 1O and 24\% of 2O Cepheids. The
period ratios of about 1 may indicate non-radial pulsations, but
it may also be an indicator of Cepheids with high rates of period
changes. Poleski (2008) showed that the latter behavior is very common
among LMC Cepheids, in the SMC it may be even more frequent.

\Section{The Catalog of Classical Cepheids in the SMC}
The OGLE-III catalog of classical Cepheids in the SMC contains 4630 stars:
2626 fundamental-mode, 1644 first-overtone, 83 second-overtone, 59
double-mode F/1O, 215 double-mode 1O/2O, two triple-mode F/1O/2O and one
triple-mode 1O/2O/3O Cepheids. The catalog data are available through the
user-friendly\linebreak WWW interface or by the anonymous FTP site:

\begin{center}
{\it http://ogle.astrouw.edu.pl/} \\
{\it ftp://ftp.astrouw.edu.pl/ogle/ogle3/OIII-CVS/smc/cep/}\\
\end{center}

The FTP site is organized as follows. The file named {\sf ident.dat}
contains the full list of Cepheids. The variables have been numbered in
order of increasing right ascension and designated as OGLE-SMC-CEP-NNNN,
where NNNN is a four-digit number. The consecutive columns of the {\sf
ident.dat} file contain: the object designation, OGLE-III field and
internal database number of a star (consistent with the photometric maps of
the SMC by Udalski \etal 2008b), mode(s) of pulsation (F, 1O, 2O, F/1O,
1O/2O, F/1O/2O, or 1O/2O/3O), equinox J2000.0 right ascension and
declination, cross-identifications with the OGLE-II catalog of Cepheids in
the SMC (Udalski \etal 1999abd), cross-identifications with the
extragalactic part of the GCVS (Artyukhina \etal 1995), and other
designations taken from the GCVS. Stars without the internal OGLE-III
number provided have no counterparts in the SMC photometric maps. These are
usually bright saturated objects, only with the {\sc DoPhot} photometry
available.

\begin{figure}[htb]
\centerline{\includegraphics[width=13.2cm]{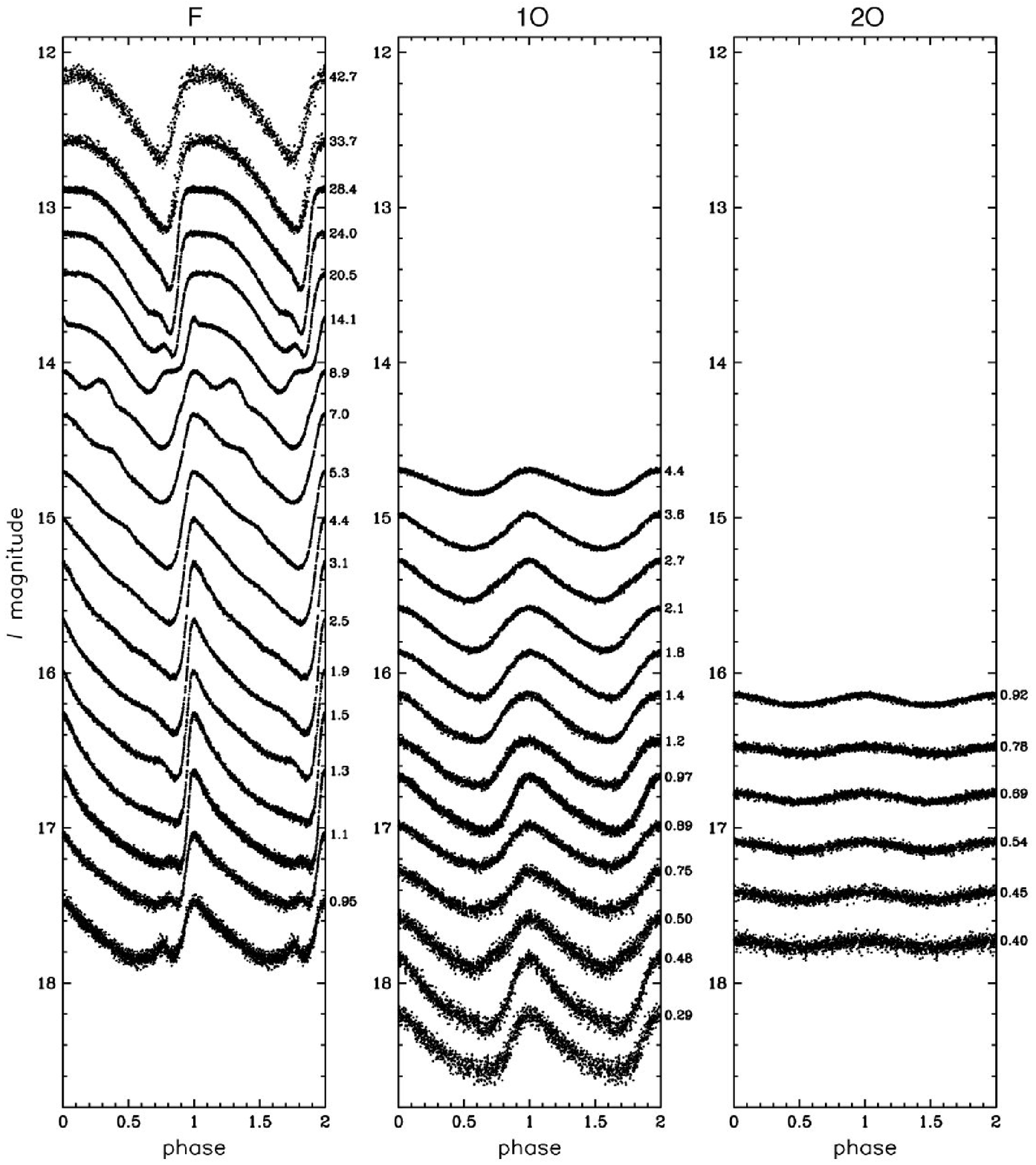}}
\FigCap{Examples of light curves of single-mode Cepheids in the SMC. {\it
Left panel} presents fundamental-mode stars, {\it middle panel} --
first-overtone, and {\it right panel} -- second-overtone pulsators. Small
numbers at the right side of each panel show the rounded periods in days of
the light curves presented in panels.}
\end{figure}

\hglue-6pt
Files {\sf cepF.dat}, {\sf cep1O.dat}, {\sf cep2O.dat}, {\sf cepF1O.dat},
{\sf cep1O2O.dat}, {\sf cepF1O2O.dat}, and {\sf cep1O2O3O.dat} contain
observational properties of the single-, double- and triple-mode Cepheids
with the corresponding modes excited. We provide there: the intensity mean
magnitudes in the {\it I}- and {\it V}-bands, periods in days and their
uncertainties (derived with the {\sc Tatry} code of Schwarzenberg-Czerny
1996), epochs of maximum light, peak-to-peak amplitudes in the {\it
I}-band, and Fourier parameters $R_{21}$, $\phi_{21}$, $R_{31}$,
$\phi_{31}$ derived for {\it I}-band light curves.

The {\sf remarks.txt} file contains additional information on some
Cepheids: comments about uncertain classifications, high rates of period
change, blends, additional periods, etc. The subdirectory {\sf phot/}
contains multi-epoch {\it I}- and {\it V}-band OGLE photometry of the
stars. If available, the OGLE-II data have been merged with the OGLE-III
photometry. The subdirectory {\sf fcharts/} contains finding charts of all
objects. These are the $60\arcs\times60\arcs$ subframes of the {\it I}-band
DIA reference images, oriented with N up, and E to the left.

Fig.~7 illustrates typical {\it I}-band light curves of single-mode
Cepheids in our sample. Two of the brightest light curves were obtained
with the {\sc DoPhot} package on partly saturated stars, therefore this
photometry is more scattered than the regular OGLE data. The Hertzsprung
progression (Hertzsprung 1926) of the light curve shapes is clearly visible
for F Cepheids.

\begin{figure}[htb]
\centerline{\includegraphics[width=12.7cm]{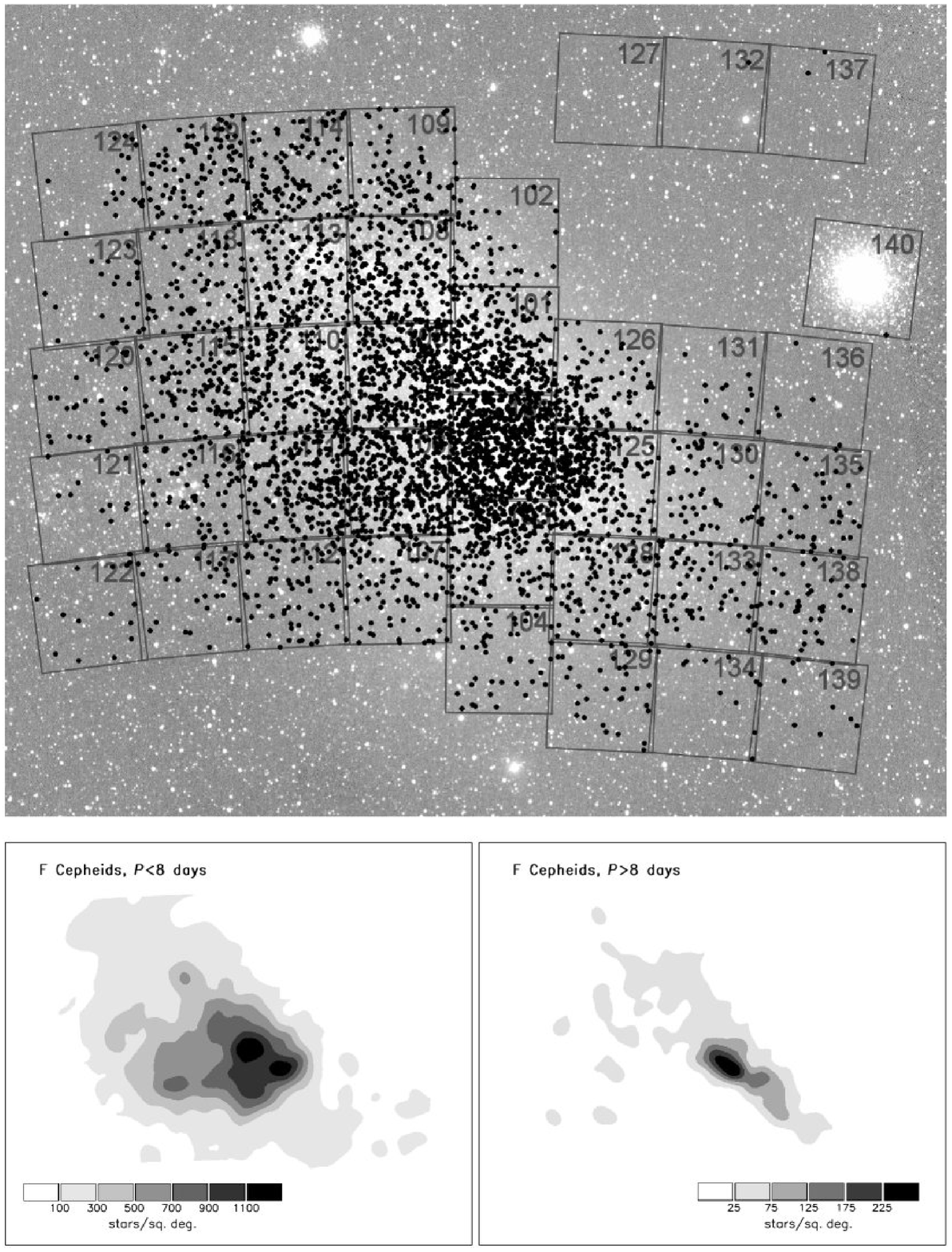}}
\FigCap{Spatial distribution of classical Cepheids in the SMC. The
background image in the {\it upper panel} is originated from the ASAS wide
field sky survey (Pojmañski 1997). The contours show OGLE-III fields. {\it
Bottom panels} show the distribution of fundamental-mode Cepheids with
$P<8$~days ({\it left panel}) and $P>8$~days ({\it right panel}).}
\end{figure}

The spatial distribution of classical Cepheids in the SMC is presented in
Fig.~8. Bottom panels show surface density maps for fundamental-mode
Cepheids with periods shorter (left panel) and longer (right panel) than
8~days. The distribution of fundamental-mode Cepheids is highly correlated
with their periods. The longer-period Cepheids are concentrated toward the
bar of the SMC, while the shorter-period pulsators are distributed more
homogeneously over the galaxy.

As a test of completeness we cross-identified the preliminary version of
our catalog with the largest sample of Cepheids in the SMC published to
date -- the OGLE-II catalog of Cepheids (Udalski \etal 1999abd). We did not
find counterparts for 19 objects from the OGLE-II list. Seven of the
missing Cepheids were saturated in the OGLE-III DIA database, and in the
final version of the OIII-CVS we included the {\sc DoPhot} photometry for
these stars. Further ten objects have been reclassified as non-Cepheid
variables (usually as ellipsoidal variables). Only two missing stars
occurred to be unsaturated classical Cepheids, both were affected by a poor
quality of the photometry due to the neighborhood of a very bright
star. The missing Cepheids were added to the final version of our catalog.

We also compared our list of double-mode Cepheids with the EROS-2 catalog
of 41 F/1O and 129 1O/2O Cepheids by Marquette \etal (2009). Four EROS-2
double-mode Cepheids lie outside the OGLE-III fields. Our catalog contains
all the remaining stars, but three of them were not classified as
double-mode Cepheids. One objects was recognized in our catalog as a
superposition of two Cepheids (double Cepheid), in two other cases we could
not identify the secondary mode in the OGLE data. The tests show that our
catalog of classical Cepheids in the SMC is practically complete in the
regions covered by the OGLE-III fields.

\Section{Comparison of the SMC and LMC Cepheids}
The large and complete samples of variable stars presented in the OIII-CVS
give an opportunity to compare various stellar populations in the LMC, SMC
and, in the future, in the Galaxy. Fig.~9 shows period distributions of
\begin{figure}[htb]
\centerline{\includegraphics[width=12.3cm, bb=45 215 565 755]{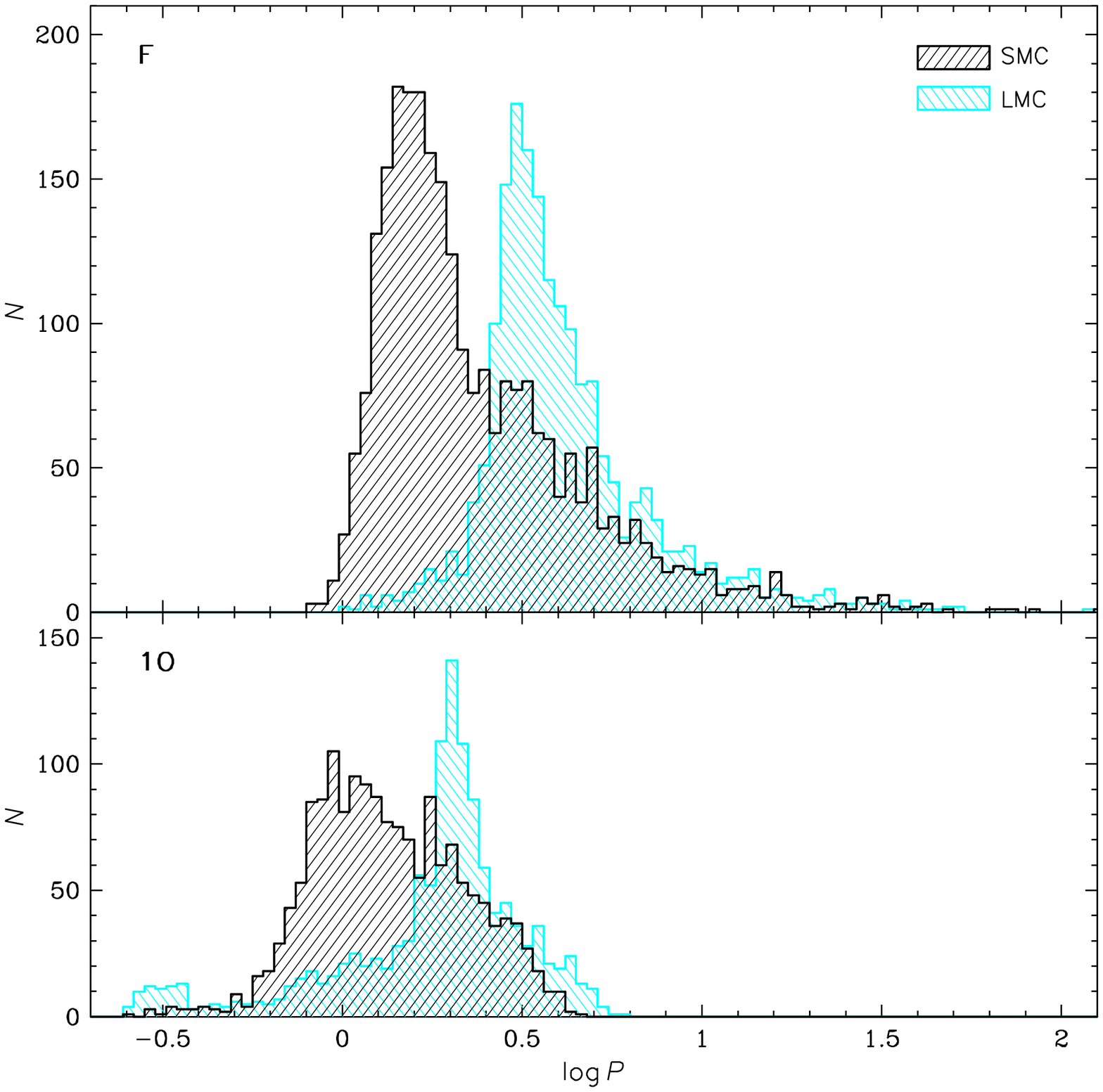}}
\FigCap{Period distribution of single-mode Cepheids in the SMC (black) and
LMC (cyan). {\it Upper panel} shows fundamental-mode, and {\it bottom panel}
-- first-overtone Cepheids.}
\end{figure}
single-mode Cepheids in both Magellanic Clouds, separately for the
fundamental mode and first overtone. In Fig.~10 we compare the {\it I}-band
apparent mean magnitude distributions of the same samples. There is no
doubt that LMC and SMC host two different populations of classical
Cepheids. In the SMC there are many more short-period and fainter Cepheids
in each pulsation mode. This difference in the period distribution may be
explained by different metal abundances in both galaxies (Hofmeister 1967,
Becker \etal 1977). The low-massive metal-poor stars experience more
extended blue loop excursion during their evolution in the HR
diagram. Thus, in the metal-poor environments (like in the SMC) the
low-mass cut-off for the Cepheids crossing the instability strip for the second
and the third time is lower than in the more metal-rich galaxies (like in
the LMC).
\begin{figure}[htb]
\centerline{\includegraphics[width=12.3cm, bb=45 215 565 755]{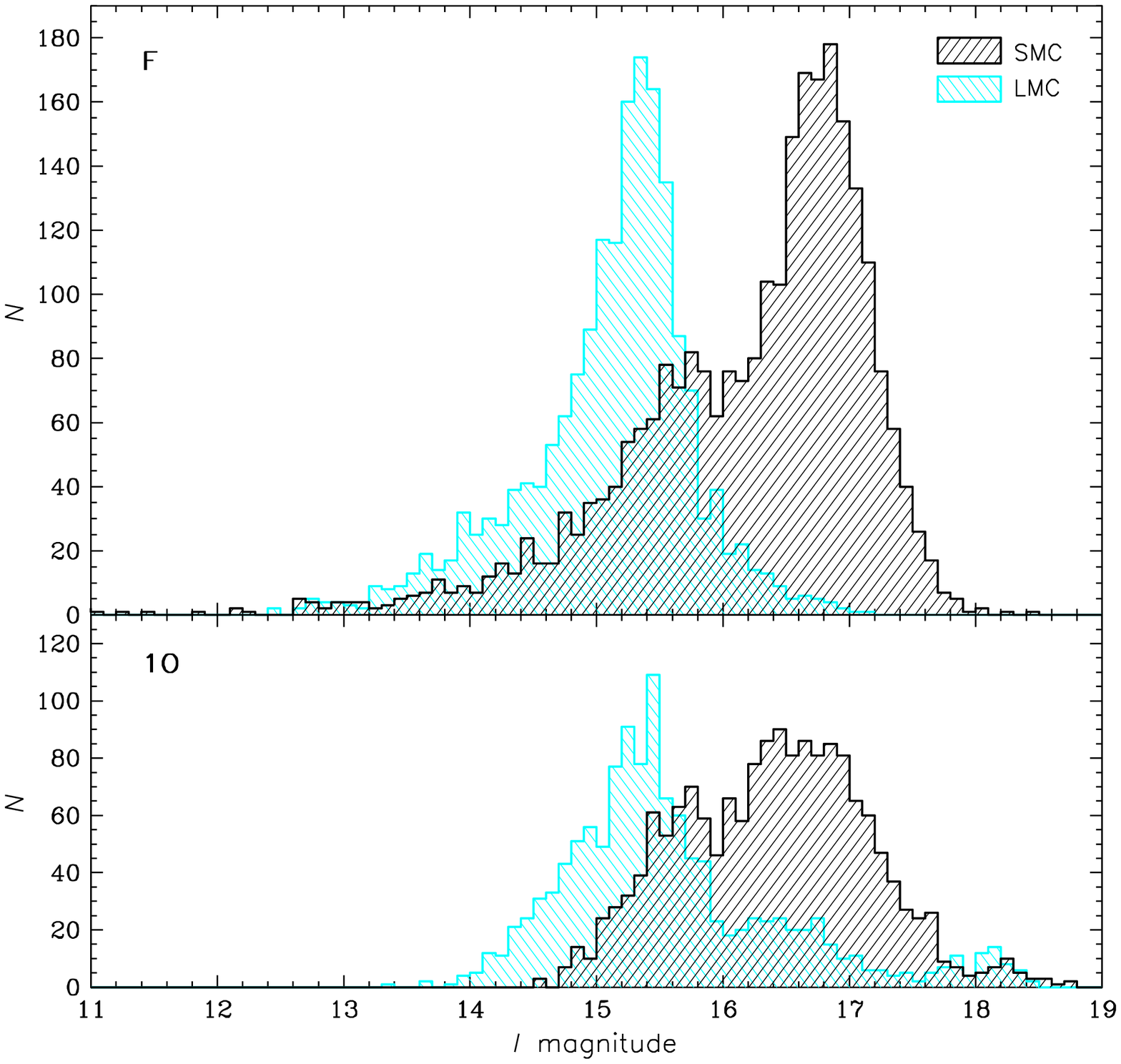}}
\vskip-6pt
\FigCap{Distribution of mean {\it I}-band magnitudes of single-mode
Cepheids in the SMC (black) and LMC (cyan). {\it Upper panel} shows
fundamental-mode, and {\it bottom panel} -- first-overtone Cepheids.}
\end{figure}

However, the period and luminosity distributions of fundamental-mode
Cepheids in the SMC seem to have two maxima: the main peak is at
$\log{P}\approx0.2$ ($I\approx16.8$~mag) and the secondary maximum is at
$\log{P}\approx0.5$ ($I\approx15.8$~mag). The latter maximum is the same as
the main frequency peak of the LMC Cepheids (for magnitudes one has to
apply appropriate shift of about 0.4~mag to compensate different distances
and reddenings toward both Clouds). The first-overtone pulsators seem to
have three maxima in the period and luminosity distributions in both
galaxies: at $\log{P}\approx-0.5$, $\log{P}\approx0.0$, and
$\log{P}\approx0.3$. The difference between LMC and SMC lays in different
numbers of Cepheids in every group.

We conclude that each of the Magellanic Clouds hosts two or three different
populations of classical Cepheids, which differ in periods, luminosities
and also probably in metal abundances and ages. The spatial distribution of
Cepheids in the SMC, which is different for short- and long-period
Cepheids, confirms this conclusion. In both galaxies we can find
representatives of each population of Cepheids, but the LMC prefers
the long-period pulsators, while in the SMC the short-period Cepheids
dominate.

\Section{Period--Luminosity Relation}
The PL relations for classical Cepheids are one of the most powerful tools
to measure distances within and outside the Galaxy. The Magellanic Clouds
play an unique role in the calibrations of PL relations, because both
galaxies are at small distances to us, have different metal abundance, and
host very rich populations of Cepheids.

\begin{figure}[p]
\vspace{-0.8cm}
\centerline{\includegraphics[width=13.0cm]{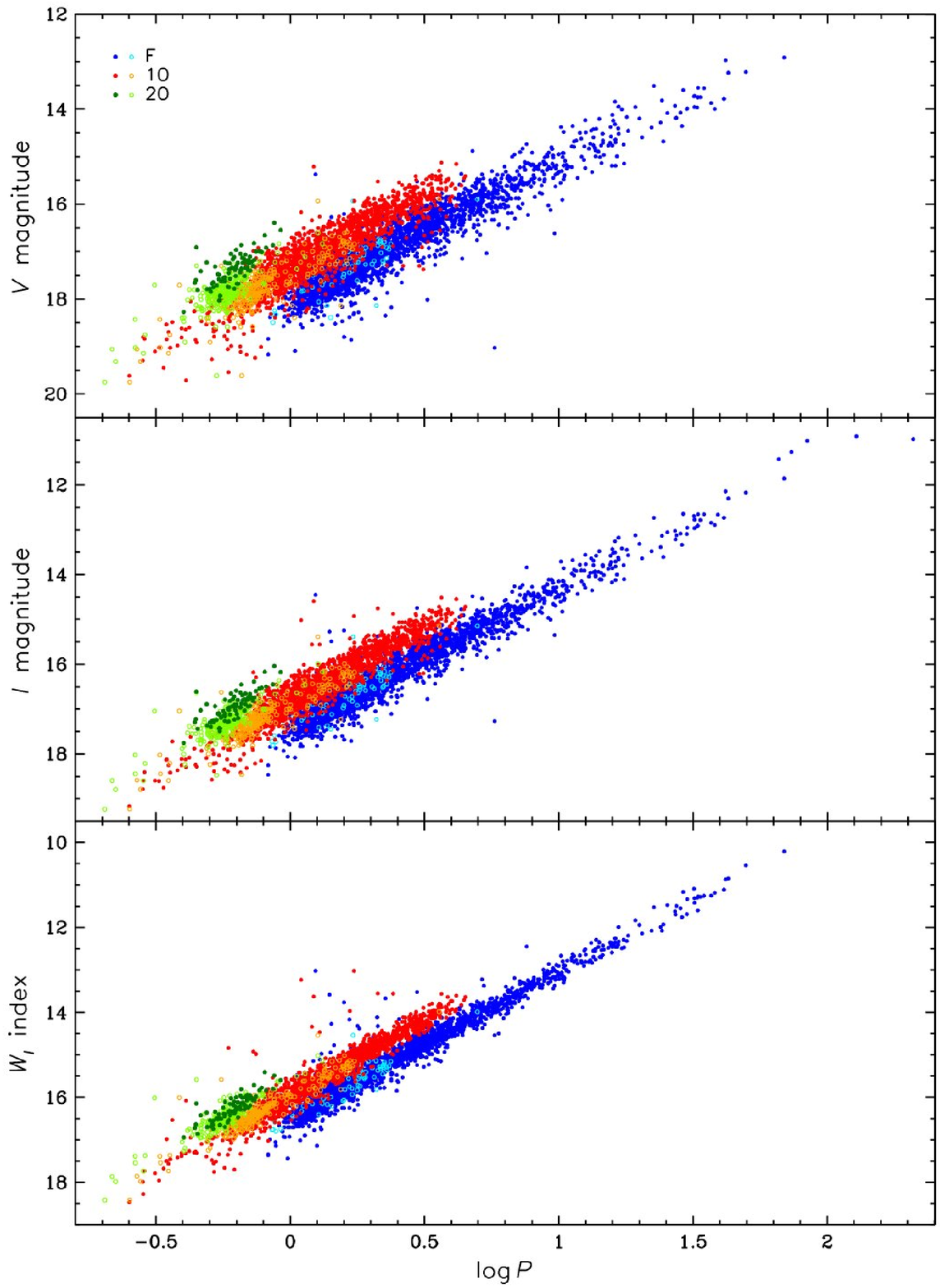}}
\vspace{-0.8cm}
\FigCap{Period--luminosity relations for classical Cepheids in the SMC. The
color symbols are the same as in Fig.~2.}
\end{figure}

The PL diagrams in the {\it V}- and {\it I}-band magnitudes and in the
$W_I$ Wesenheit index are shown in Fig.~11. The scatter of the relations is
generally larger than in the LMC (Paper~I), because the SMC is thought to
have larger depth along the line of sight. However, in the LMC we
observed more individual Cepheids located significantly below the
$\log{P}$--$I$ and $\log{P}$--$V$ relations, \ie objects suffering from
heavy reddening. It shows that the spatial distribution of interstellar
matter in the SMC is more homogeneous than in the LMC. The points located
above the appropriate PL relations are usually affected by blending.

The fundamental-mode Cepheids show a break in the slope of the PL
relation for a period of about 2~days. This non-linearity was first
reported by Bauer \etal (1999) and confirmed by Udalski \etal (1999c),
Sharpee \etal (2002) and Sandage \etal (2009). This feature may be another
clue that the SMC hosts at least two populations of classical Cepheids --
stars with shorter periods have a different chemical composition and thus
obey a different PL law.

The non-linear relation between $\log{P}$ and luminosities of Cepheids in
the SMC complicates the efforts to compare PL laws in various
environments. Following Udalski \etal (1999c) and Sandage \etal (2009) we
removed from our solution all F Cepheids with $\log{P}<0.4$. We also
excluded from the sample several of the longest-period Cepheids that
saturate on the OGLE frames. We take no account for possible break of the
linearity of the PL relation at $P=10$~days for F Cepheids (Sandage \etal
2009). The PL relations provided below are not compensated for interstellar
extinction, so for the $\log{P}$--$I$ and $\log{P}$--$V$ relations we can
compare the slopes of the relations only.

The least squares solution with $3\sigma$ clipping gives the following PL
relations for single-mode fundamental-mode classical Cepheids in the SMC:
\begin{eqnarray*}
V  &=&-2.646(\pm0.036)\log{P}+17.792(\pm0.026)\qquad\sigma=0.28~{\rm mag}\\
I  &=&-2.908(\pm0.029)\log{P}+17.240(\pm0.021)\qquad\sigma=0.22~{\rm mag}\\
W_I&=&-3.326(\pm0.019)\log{P}+16.383(\pm0.014)\qquad\sigma=0.15~{\rm mag}
\end{eqnarray*}
and for first-overtone Cepheids:
\begin{eqnarray*}
V  &=&-3.171(\pm0.034)\log{P}+17.372(\pm0.008)\qquad\sigma=0.28~{\rm mag}\\
I  &=&-3.349(\pm0.027)\log{P}+16.827(\pm0.007)\qquad\sigma=0.23~{\rm mag}\\
W_I&=&-3.623(\pm0.020)\log{P}+15.971(\pm0.005)\qquad\sigma=0.16~{\rm mag.}
\end{eqnarray*}

The slopes of the PL relations agree within up to $2\sigma$ with the fits
of Udalski \etal (1999c)\footnote{the revised coefficients of the fit are
available from:\\ {\it
ftp://ftp.astrouw.edu.pl/ogle/ogle2/var\_stars/smc/cep/catalog/README.PL}}
to the OGLE-II sample of Cepheids in the SMC (with exception of the
$\log{P}$--$W_I$ relation for 1O Cepheids, where the difference is at the
level of $2.4\sigma$). It would be interesting to compare the PL relations
in both Magellanic Clouds to answer the question how different metal
abundances influence the PL laws of Cepheids. Recently, Sandage \etal
(2009) stated that there is a statistically significant difference between
slopes of the PL relations for LMC and SMC Cepheids. Indeed, comparing the
above fits with the PL laws obeyed by the LMC fundamental-mode Cepheids
(Paper~I, Ngeow \etal 2009) we find a difference in the slopes of the
$\log{P}$--$V$ relations at the level $>3\sigma$. In the {\it I}-band the
difference has a significance of only $1.8\sigma$, while the
$\log{P}$--$W_I$ relations for F Cepheids are statistically
indistinguishable in both Clouds. For the first-overtone Cepheids the {\it
V}-band PL relations in the LMC and SMC are almost the same, the {\it
I}-band relations differ at the level of about $2\sigma$, while the
$\log{P}$--$W_I$ relations differ very significantly, at the level of
$>8\sigma$.

\Section{Conclusions}
In this paper we described the catalog of 4630 classical Cepheids in the
SMC. This is the largest set of Cepheids detected so far in any
environment. We make available to the astronomical community a precise,
standard {\it VI} photometry of these stars collected within 8 (OGLE-III)
or 13 (OGLE-II + OGLE-III) years of observations. These data may be
utilized for studying many fundamental problems, such as the universality
of the PL and PLC relations, the accuracy of stellar modeling, the
structure and history of the SMC. With the OGLE-III catalog of 3375
classical Cepheids in the LMC (Paper~I) our samples provide a unique
opportunity to perform detailed comparative studies between both
populations.

Among large number of variable stars one can discover very rare objects and
phenomena. Our preliminary analysis described in this paper revealed
extraordinarily large set of single-mode second-overtone Cepheids, three
triple-mode pulsators, two classical Cepheids with eclipses, and five
double Cepheids. We discovered a large number 1O Cepheids with additional
periods of about 0.60--0.65 of the primary ones.

\Acknow{We are grateful to W.~Dziembowski and S.~Sheppard for a critical
reading of the manuscript. We thank Z.~Ko³aczkowski, G.~Pojmañski,
A.~Schwar\-zenberg-Czerny and J.~Skowron for providing the software and
data which enabled us to prepare this study.

This work has been supported by MNiSW grants: NN203293533 to IS and
N20303032/4275 to AU. The massive period search was performed at the
Interdisciplinary Centre for Mathematical and Computational Modeling of
Warsaw University (ICM), project no.~G32-3. We wish to thank M. Cytowski
for his skilled support.}


\begin{references}
\refitem{Alard, C., and Lupton, R.H.}{1998}{\ApJ}{503}{325}
\refitem{Alard, C.}{2000}{\AAS}{144}{363}
\refitem{Alcock, C., \etal (MACHO collaboration)}{1997}{arXiv:astro-ph/9709025}{~}{~}
\refitem{Antonello, E., and Kanbur, S.M.}{1997}{\MNRAS}{286}{L33}
\refitem{Artyukhina, N.M., \etal}{1995}{~}{~}{General Catalogue of Variable Stars, 4rd ed., vol.V. Extragalactic Variable Stars, "Kosmosinform", Moscow}
\refitem{Bauer, F., \etal (EROS collaboration)}{1999}{\AA}{348}{175}
\refitem{Beaulieu, J.P., \etal (EROS collaboration)}{1997}{\AA}{321}{L5}
\refitem{Becker, S.A., Iben, I., Jr., and Tuggle, R.S.}{1977}{\ApJ}{218}{633}
\refitem{Bernard,~E.J., Monelli,~M., Gallart,~C., Aparicio,~A., Cassisi,~S., Drozdovsky,~I., Hidalgo,~S.L., Skillman,~E.D., and Stetson,~P.B.}{2010}{\ApJ}{712}{1259}
\refitem{Buchler, J.R., and Szab{\'o}, R.}{2007}{\ApJ}{660}{723}
\refitem{Dziembowski, W.A., and Smolec, R.}{2009a}{\Acta}{59}{19}
\refitem{Dziembowski, W.A., and Smolec, R.}{2009b}{~}{~}{in ``Stellar Pulsation: Challenges for Theory and Observation'', Eds. J.A.~Guzik and P.A.~Bradley, {\it AIP Conf. Proc.}, 1170, 83}
\refitem{Hertzsprung, E.}{1926}{Bull. Astr. Inst. Netherlands}{3}{115}
\refitem{Hofmeister, E.}{1967}{Z. Astrophys.}{65}{194}
\refitem{Kato, D., \etal}{2007}{PASJ}{59}{615}
\refitem{Leavitt, H.S.}{1908}{Ann. Harv. Coll. Obs.}{60}{87}
\refitem{Marquette, J.B., \etal}{2009}{\AA}{495}{249}
\refitem{Moskalik, P., and Ko{\l}aczkowski, Z.}{2008}{Comm. in Asteroseismology}{157}{343}
\refitem{Ngeow, C.-C., Kanbur, S.M., Neilson, H.R., Nanthakumar, A., and Buonaccorsi, J.}{2009}{\ApJ}{693}{691}
\refitem{Payne-Gaposchkin, C., and Gaposchkin, S.}{1966}{Smithsonian Contrib. Astrophys.}{9}{1}
\refitem{Pojmañski, G.}{1997}{\Acta}{47}{467}
\refitem{Poleski, R.}{2008}{\Acta}{58}{313}
\refitem{Poleski,~R., Soszyñski,~I., Udalski, A., Szymañski,~M.K., Kubiak,~M., Pietrzyñski,~G., Wyrzykowski,~£., Szewczyk,~O., and Ulaczyk,~K.}{2010}{\Acta}{60}{1}
\refitem{Sandage, A., Tammann, G.A., and Reindl, B.}{2009}{\AA}{493}{471}
\refitem{Schechter, P.L., Mateo, M., and Saha, A.}{1993}{\PASP}{105}{1342}
\refitem{Schwarzenberg-Czerny, A.}{1996}{\ApJ}{460}{L107}
\refitem{Shapley, H., and McKibben Nail, V.}{1955}{Proc. Nat. Acad. Sci.}{41}{829}
\refitem{Sharpee, B., Stark, M., Pritzl, B., Smith, H., Silbermann, N., Wilhelm, R., and Walker, A.}{2002}{\AJ}{123}{3216}
\refitem{Simon, N.R., and Lee, A.S.}{1981}{\ApJ}{248}{291}
\refitem{Soszyñski,~I., Poleski,~R., Udalski, A., Kubiak,~M., Szymañski,~M.K., Pietrzyñski,~G., Wyrzykowski,~£., Szewczyk,~O., and Ulaczyk,~K.}{2008a}{\Acta}{58}{153}
\refitem{Soszyñski,~I., Poleski,~R., Udalski, A., Szymañski,~M.K., Kubiak,~M., Pietrzyñski,~G., Wyrzykowski,~£., Szewczyk,~O., and Ulaczyk,~K.}{2008b}{\Acta}{58}{163 (Paper~I)}
\refitem{Soszyñski,~I., Udalski, A., Szymañski,~M.K., Kubiak,~M., Pietrzyñski,~G., Wyrzykowski,~£., Szewczyk,~O., Ulaczyk,~K., and Poleski,~R.}{2008c}{\Acta}{58}{293}
\refitem{Soszyñski,~I., Udalski, A., Szymañski,~M.K., Kubiak,~M., Pietrzyñski,~G., Wyrzykowski,~£., Szewczyk,~O., Ulaczyk,~K., and Poleski,~R.}{2009a}{\Acta}{59}{1}
\refitem{Soszyñski,~I., Udalski, A., Szymañski,~M.K., Kubiak,~M., Pietrzyñski,~G., Wyrzykowski,~£., Szewczyk,~O., Ulaczyk,~K., and Poleski,~R.}{2009b}{\Acta}{59}{239}
\refitem{Szymañski, M.K.}{2005}{\Acta}{55}{43}
\refitem{Udalski, A.}{2003}{\Acta}{53}{291}
\refitem{Udalski, A., Soszyñski, I., Szymañski, M., Kubiak, M., Pietrzyñski, G., Wo¼niak, P., and ¯ebruñ, K.}{1999a}{\Acta}{49}{1}
\refitem{Udalski, A., Soszyñski, I., Szymañski, M., Kubiak, M., Pietrzyñski, G., Wo¼niak, P., and ¯ebruñ, K.}{1999b}{\Acta}{49}{45}
\refitem{Udalski, A., Szymañski, M., Kubiak, M., Pietrzyñski, G., Soszyñski, I., Wo¼niak, P., and ¯ebruñ, K.}{1999c}{\Acta}{49}{201}
\refitem{Udalski, A., Soszyñski, I., Szymañski, M., Kubiak, M., Pietrzyñski, G., Wo¼niak, P., and ¯ebruñ, K.}{1999d}{\Acta}{49}{437}
\refitem{Udalski, A., Szymañski, M.K., Soszyñski, I., and Poleski. R.}{2008a}{\Acta}{58}{69} 
\refitem{Udalski, A., Soszyñski, I., Szymañski, M.K., Kubiak, M., Pietrzyñski, G., Wyrzykowski, £., Szewczyk, O., Ulaczyk, K., and Poleski, R.}{2008b}{\Acta}{58}{329}
\refitem{Wo¼niak, P.R.}{2000}{\Acta}{50}{421}
\end{references}
\end{document}